\documentclass[a4paper,11pt]{article}
\pdfoutput=1 
\usepackage{jinstpub} 

\title{\boldmath Science with hot astrophysical plasmas}

\author[a,b,1]{J.S. Kaastra,\note{Corresponding author.}}
\author[a]{L. Gu,}
\author[a,b]{J. Mao,}
\author[a]{M. Mehdipour,}
\author[a,b]{F. Mernier,}
\author[a]{J. de Plaa,}
\author[a]{A.J.J. Raassen,}
\author[a,b]{I. Urdampilleta}

\affiliation[a]{SRON Netherlands Institute for Space Research,\\Sorbonnelaan 2, 3584 CA Utrecht, The Netherlands}
\affiliation[b]{Leiden Observatory, Leiden University,
\\PO Box 9513, 2300 RA Leiden, the Netherlands}

\emailAdd{j.kaastra@sron.nl}

\abstract{We present some recent highlights and prospects 
for the study of hot astrophysical plasmas.
Hot plasmas can be studied primarily through their X-ray emission and absorption. Most astrophysical objects,
from solar system objects to the largest scale structures of the Universe, contain hot gas. In general we can
distinguish collisionally ionised gas and photoionised gas. We introduce several examples of both classes
and show where the frontiers of this research in astrophysics can be found. We put this also in the context
of the current and future generation of X-ray spectroscopy satellites. The data coming from these missions
challenge the models that we have for the calculation of the X-ray spectra.}


\arxivnumber{1234.56789} 

\proceeding{2$^{\text{nd}}$ European Conference on Plasma Diagnostics\\
  18--21 April 2017\\
  Bordeaux}

\begin{document}
\maketitle
\flushbottom

\section{Introduction}

X-ray emitting plasmas are found everywhere in the Universe: from the 
Solar System to cosmic web filaments. They
can be found in a 
broad range of environments and for a range of physical conditions:
collisionally ionised, photo-ionised, transiently ionised.

Gravity is the ultimate energy source for cosmic plasmas. This can be direct, through the 
accretion of matter onto massive objects. The gravity leads to infall and accretion, which produces heating through the gain in potential energy; the heating of the plasma then produces  radiation.
Alternatively, nuclear fusion in collapsed objects is ultimately also caused by gravity: it leads to compression of the gas, causing nuclear fusion that constitutes the heat source producing the radiation.
Finally,  electromagnetic processes are also induced by the motions caused by gravity fields; the produced electric and magnetic field act in accelerating and heating the matter, again leading to radiation.

There are principally three methods currently in use to measure X-ray spectra in space. CCDs allow for
imaging and low/medium resolution spectroscopy.
Examples are the ASCA satellite (1993--2000), and the Chandra and XMM-Newton observatories, both launched in 1999 and still operational.

Gratings are mostly used to study point sources. They have 
poor imaging capabilities but deliver medium/high (few 100) resolution X-ray spectra.
Examples are EUVE (1992--2001), and the grating spectrometers on the Chandra and XMM-Newton satellites.

A more recent development are the
calorimeters. They offer
imaging capabilities and high spectral resolution.
Up to now, the only example is Hitomi, which operated for a few weeks in 2016.

Apart from detecting X-rays, it is also necessary to model the resulting X-ray spectra. Work on modelling astrophysical X-ray spectra started in the late 1960s. In this paper we mainly focus on the code originally developed by Rolf Mewe, whose first publication on it appeared in 1972. Since then, the code evolved gradually, initially by Mewe and his collaborators Gronenschild and others; later Kaastra joined the project. In 1991 this code evolved into the SPEX package, of which now version 3 (released January 2016) is the main working horse. 
It evolved from a pure plasma model to a full astrophysical modelling machine including data analysis (fitting), plotting and diagnostic output options. The package can be found under www.sron.nl/spex.

\section{Collisional Ionisation Equilibrium}

One very common type of plasma found in astrophysical sources are the plasmas in collisional ionisation equilibrium (CIE). Collisions with mainly thermal electrons determine the production of X-rays, and the effects of external radiation fields can be neglected. Most common here is the coronal approximation, valid for not too high densities.

Examples of such plasmas are the corona of our Sun and of cool stars in general, the winds of hotter stars, the hot ($10^6$~K) component of the interstellar medium of our galaxy, clusters of galaxies and the cosmic web that connects these clusters.
In this section we focus on clusters of galaxies.

Most of the baryonic matter of clusters is in the form of a hot ($10^{7\text{--}8}$ K), optically thin plasma, namely the intra-cluster medium (ICM). 
This plasma is also found to be in, or very close to CIE. In that way, the ICM is an immense celestial laboratory, in which the CIE plasma is confined only by the cluster gravitational potential well (essentially shaped by its dark matter component). The thermodynamical and chemical properties of the ICM, which can be probed via X-ray spectroscopy from CCDs and gratings on board current missions, provide crucial information on the formation of those large scale structures, and on the various sources of feedback that they undergo with cosmic time.

\begin{figure}[t]
\centering
\includegraphics[width=.48\textwidth]{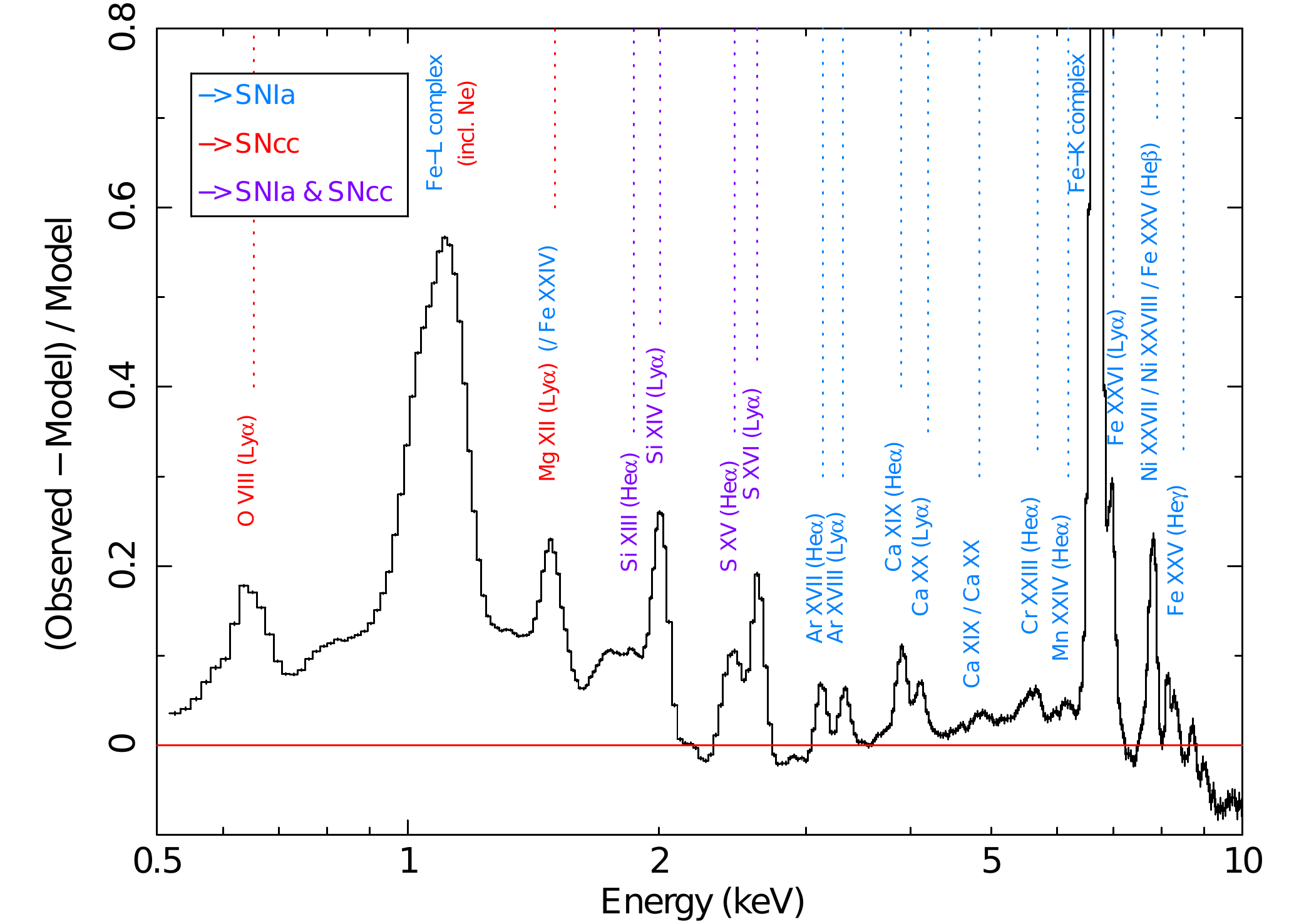}
\qquad
\includegraphics[width=.46\textwidth]{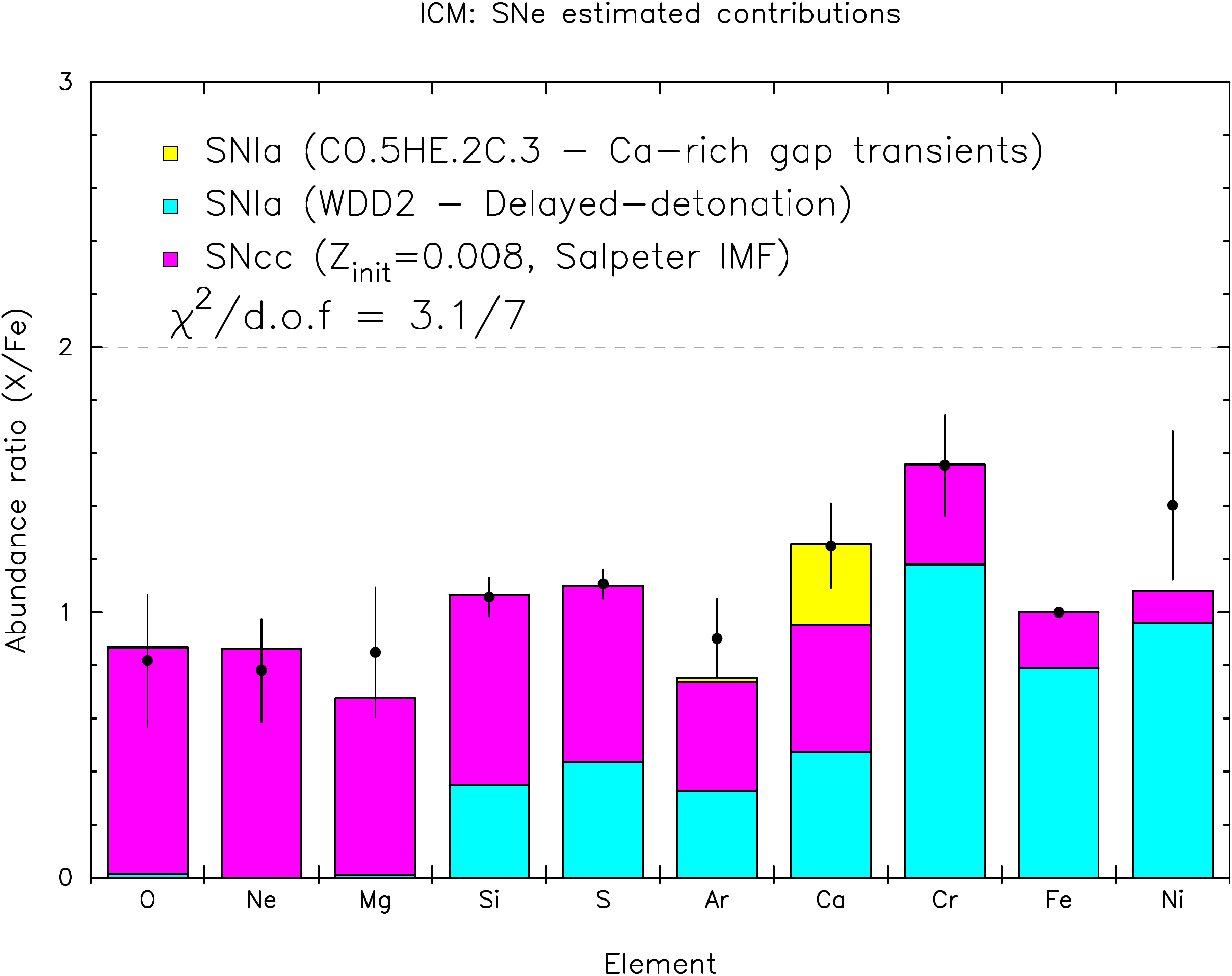}
\caption{\label{fig:CHEERS_results} \textit{Left:} Stacked X-ray spectra of a sample of 44 galaxy clusters, groups, and ellipticals, as seen by the XMM-Newton EPIC (CCD) cameras. The emission lines detected in the ICM (O, Ne, Mg, Si, S, Ar, Ca, Cr, Mn, Fe, and Ni) can be easily converted into elemental abundances \cite{mernier2016a}. \textit{Right:} Average abundance ratios (relative to Fe) of the same sample (black data points) compared to the best combination of supernova nucleosynthesis models (histograms) \cite{mernier2017}.}
\end{figure}

One good example of the significant contribution of X-ray spectroscopy to
cluster science is the emergence of the "cooling flow problem" in the past 15
years. According to the classical cooling flow model, large amounts of
intra-cluster plasma should lose most of their energy by radiation and cool down
to form hundreds of new stars per year within the core of relaxed clusters
\cite{fabian1994}. X-ray spectra of clusters were thus expected to contain
transitions of several ions emitting at very low temperatures. The instruments
on board XMM-Newton and Chandra were capable to test this paradigm, and revealed
the surprising absence of such spectral signatures, indicating that the ICM
stops cooling below $\sim$10$^7$ K \cite{peterson2003}. Nowadays, it is
generally accepted that central cooling flows are reduced by the mechanical
feedback of the active galactic nucleus (AGN) at the centre of the main giant
elliptical galaxy\cite{mcnamara2012}. However, what are the precise feedback
mechanisms responsible for efficiently reheating the core of the ICM is still
unclear and debated.

X-ray spectroscopy is also an excellent tool to investigate the chemical
enrichment of the ICM. Most of the chemical elements heavier than helium
(namely, "metals") in the Universe have been synthesised in stars and supernova
explosions. This is especially true for $\alpha$-elements and Fe-peak elements,
which are mainly produced by two different types of supernovae (core-collapse
and Type Ia supernovae, respectively). Consequently, the presence of emission
lines from oxygen ($Z=6$) to nickel ($Z=28$) in X-ray spectra of clusters
indicates that the ICM has been significantly enriched by billions of supernovae
since the major epoch of star formation in the Universe
(Fig.~\ref{fig:CHEERS_results} left) \cite{deplaa2007}. The equivalent widths of
the emission lines in X-ray spectra of clusters are proportional to the
abundance of their corresponding metals in the ICM. In turn, because they
correspond to the integral yields of core-collapse and Type Ia supernovae, these
abundances can be directly compared to a broad range supernova nucleosynthesis
models. Averaged over many cluster spectra, the total (i.e. statistical plus
systematic) uncertainties of the measured ICM abundances is often less than
10--20\% \cite{mernier2016a}. Consequently, these accurate measurements may even
help to constrain the currently competing supernova yield models, and therefore
to better understand the physics of supernovae in general
(Fig.~\ref{fig:CHEERS_results} right)
\cite{deplaa2007,mernier2016b,mernier2017}.

Of course, this approach relies on the accuracy of the current spectral codes
used to model and fit cluster spectra. It is therefore crucial to devote large
efforts on updating these codes when necessary (in particular, in preparation to
future X-ray missions). For instance, the major update of SPEX---from its
version 2 to its version 3---has a significant effect on the average
measurements of some elemental abundances in the ICM, in some cases with
differences up to a factor of 2 \cite{mernier2017}.

Driven by the need to further improve on the atomic data owing to the
high-resolution spectra that become available with the future X-ray missions as
Athena or Arcus, in January 2016 SPEX version 3.0 was released. The update
follows the original strategy of Mewe et al., which is minimizing the number of
mathematical operations and data storage to yield simple and fast but accurate
calculations. Now we have a more complete code with all the elements included up
to Zn ($Z\leq$ 30), which are astrophysically the most relevant.

The most important updates contained in SPEX v3.0 are the new atomic database
prepared by Raassen et al., the update of radiative recombination (RR)
\cite{mao2016} and collisional ionisation \cite{urdampilleta2017} processes
together with an improvement of the photo-ionisation model \cite{Mehd16}. 
Moreover, a new charge exchange model \cite{gu2016a} and non-thermal electrons
effects have been included. Some of these updates are described in the
paragraphs below.

The new approach contains hundreds of thousands of energy levels from hydrogen
to zinc obtained from the literature or calculations with the Flexible Atomic
Code (\citep[FAC,][]{gu2008}). 
 
\begin{figure*}
\resizebox{\hsize}{!}{\includegraphics[angle=0]{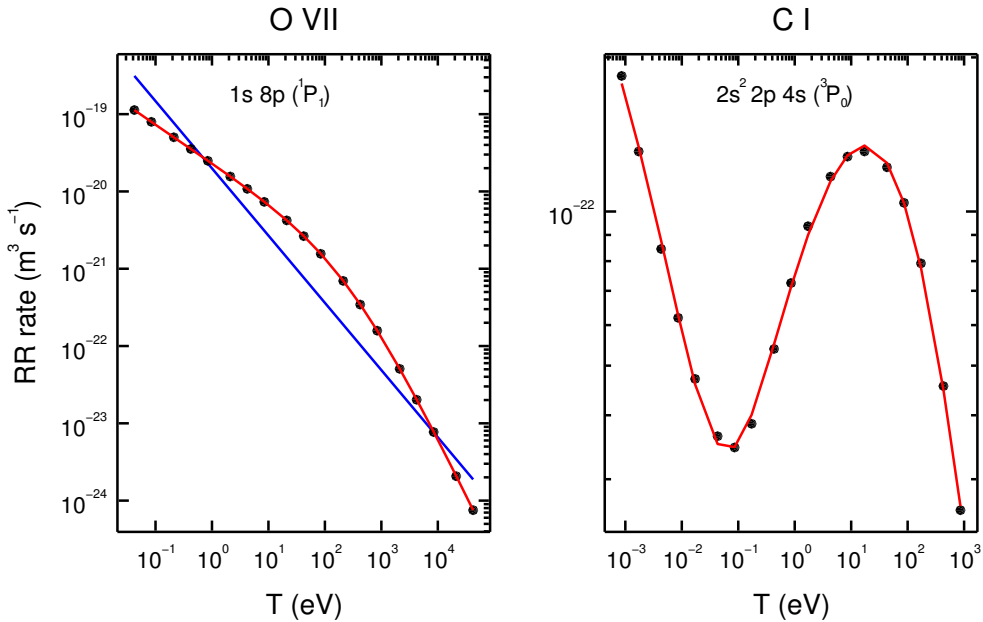}}
\caption{(left) Capture of a free electron via radiative recombination to form O VII $1s~8p~(^{1}P_{1})$. (right) Capture of a free electron to form C I $2s^2~2p~4s~(^{3}P_{0})$. Radiative recombination rates are in units of ${\rm m^{3}~s^{-1}}$). Black filled circles: ADAS calculations \cite{badnell2006}. Blue solid line: power-law implementation of the rate \cite{mewe1980}. Red solid lines: parameterisation of the ADAS results \cite{mao2016}. }
\label{fig:rr}
\end{figure*}

In previous versions of SPEX, RR rates were approximated by power
laws\cite{mewe1980}. Recombination to form H-like to Be-like ions was included
in the calculations. Such a simple implementation accounts for the general
behavior of the RR rate, namely, the rapid decrease of the rates with increasing
plasma temperature (Fig.~\ref{fig:rr} left). Nonetheless, for some levels in a
low-charge many-electron ion or atom, the RR rates do not follow the above
mentioned monotonically decreasing behavior (Fig.~\ref{fig:rr} right). 
Accordingly, we updated the SPEX code with the parameterisation of the most
recent calculations \cite{mao2016}.

In recent years, new laboratory measurements and theoretical calculations of
collisional ionisation cross-sections have become available. In SPEX v3.0 we
have updated and extended previous published compilations by introducing the
most recent experimental measurements \cite{urdampilleta2017}. Since we want to
use the rates not only for equilibrium plasmas but also for non-equilibrium
situations, it is important to know the contributions from different atomic
subshells separately.

All these updates of our spectral code were made in order to be ready for the
launch of Hitomi on 17 February 2016. This Japanese satellite carried a soft
gamma-ray detector, two hard X-ray imagers, a soft X-ray imager (CCDs) but most
importantly a unique calorimeter detector, provided by NASA. The detector is a
6x6 pixel array with 5~eV spectral resolution over the 0.5--12~keV band.

Unfortunately, the satellite was lost on March 26, 2016 due to failure of the
attitude control system. However,  during the checkout phase, one excellent
X-ray spectrum was obtained, namely from the Perseus cluster of galaxies.

First results of this mission were reported \cite{hitomi-nature}. The spectral
resolution in the Fe-K band of 5~eV is 30--40 times better than what was
available before with CCD detectors. This allowed, for instance, to fully
resolve the four strong $n=1-2$ transitions of Fe~XXV to be spectroscopically
separated, and not only that, but also to measure directly the turbulent
broadening (in addition to the thermal broadening caused by the motion of the
ions). The amount of turbulent broadening appeared to be relatively small. The
(Gaussian) turbulent velocity dispersion is only 164~km\,s$^{-1}$, and
constitutes only 4\% of the thermal pressure of the hot cluster gas. Therefore
total cluster masses determined from hydrostatic equilibrium in the central
regions need little correction for turbulent pressure.

The Hitomi spectrum offers an unprecedented benchmark of the atomic modelling of
collisional plasmas, revealing a number of places where pre-launch atomic codes
and databases such as AtomDB and SPEX need to be updated. For instance, the
AtomDB team has recently improved the wavelengths for high $n$ transitions of H-
and He-like ions, and for valence shell transitions of Li-like ions, as well as
the collisional excitation rates for H- and He-like ions. The SPEX team also
fixed a bug in the calculation of trielectronic recombination for Li-like ions,
and introduced the proper branching ratios for excitation and inner-shell
ionization to excited levels that can auto-ionize. These post-launch updates are
not made to ``fit'' the Hitomi data, but instead to reflect the needs of
analyzing a Hitomi-level spectrum.

\begin{figure}[!htbp]
\centering
\includegraphics[width=.48\textwidth]{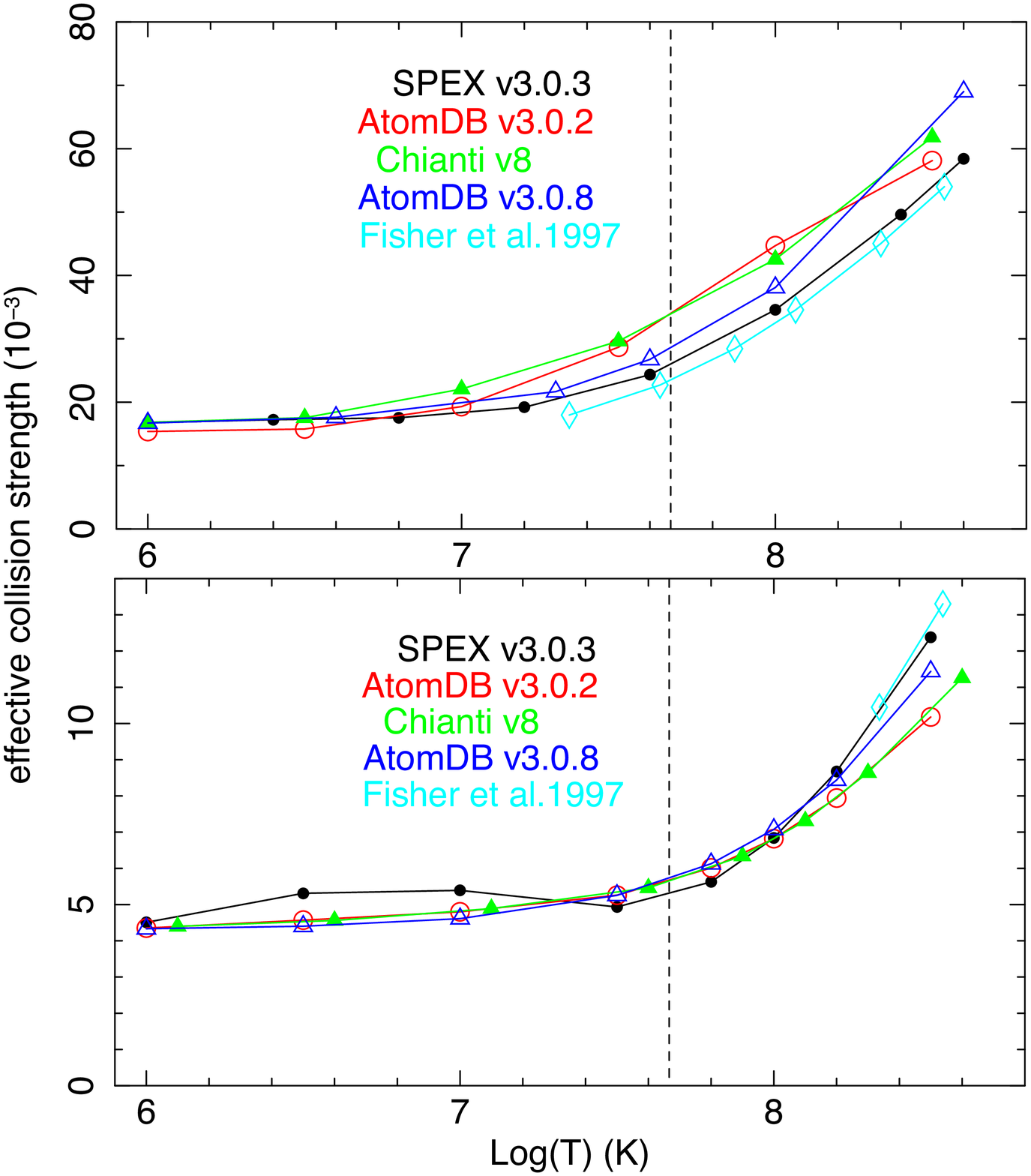}
\qquad
\includegraphics[width=.46\textwidth]{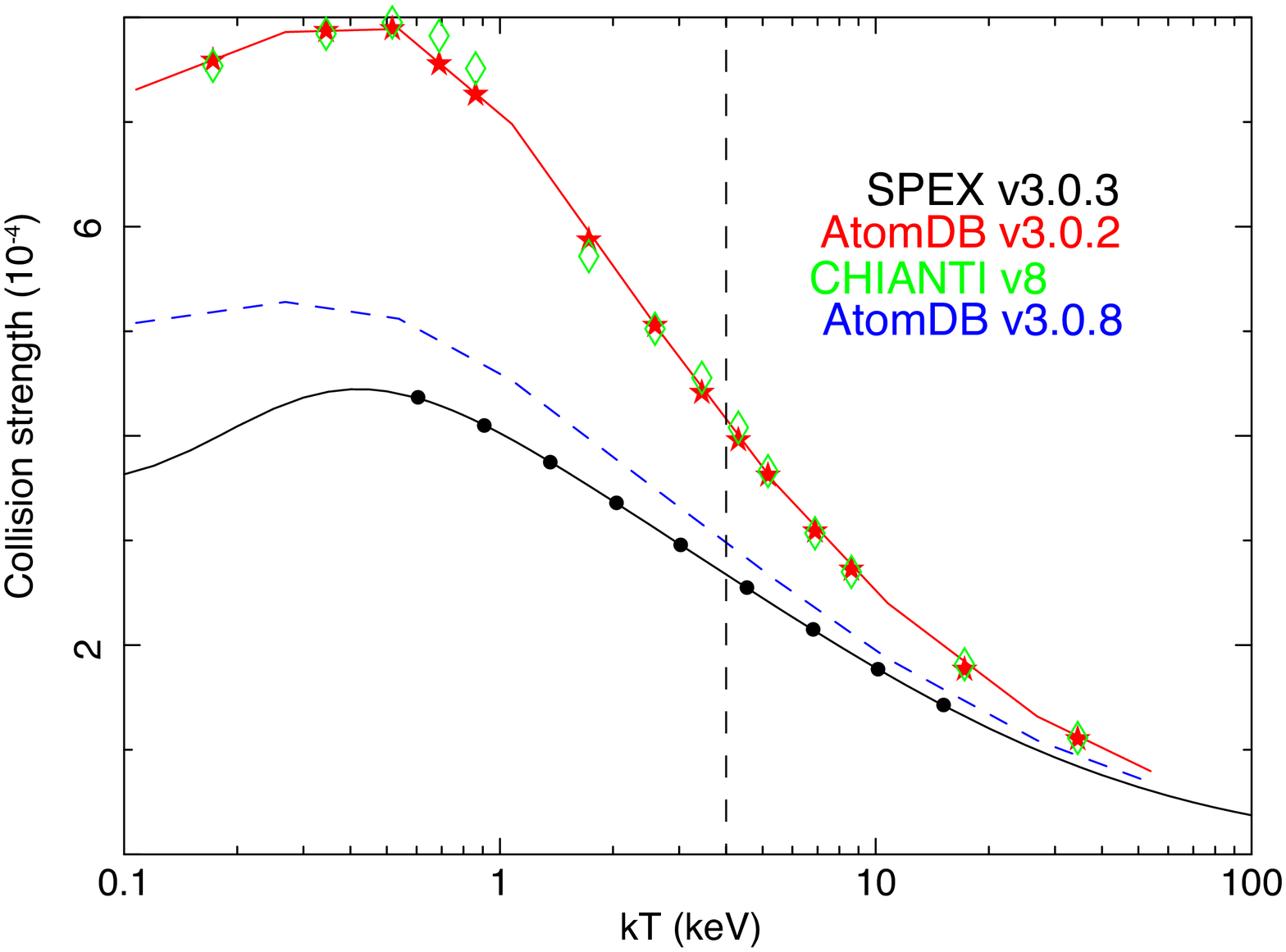}
\caption{(left) Comparisons of effective collision strength as a function of balance temperature. The Ly$\alpha$1 and Ly$\alpha$2
transitions are combined. The vertical dashed lines mark a temperature of 4~keV. (right) Comparison of the effective collision strength 
from the ground to 1s.2s ($^{3}$S$_1$) state. The vertical
line shows a temperature of 4~keV. }
\label{fig:hitomi_lya}
\end{figure}

The Hitomi analysis of the Perseus cluster also shows the dependencies of the
measurements of several astrophysically interesting parameters, such as the ICM
temperature and metal abundances, on a range of factors including instrumental
calibration, spectral fitting techniques, astrophysical modeling, and atomic
code and database.  The uncertainties of collisional excitation from the ground
state to excited levels lead to 8\%$-$16\% uncertainties for the abundances of
Si, S, Ca, and Fe, which are 2$-$17 times higher than the statistical
uncertainties of these elements. As shown in Fig.~\ref{fig:hitomi_lya}, the
effective collision strengths of ground to 2p levels for H-like Si and Fe at
4~keV (the approximate temperature of the Perseus cluster) differ by 10\%-30\%
for different (versions) of atomic codes. The differences reflect the current
state of theoretical calculations: SPEX utilizes a R-matrix calculation by
\cite{aggarwal1992}, while the AtomDB code uses a calculation based on the
distorted wave approximation by \cite{li2015}. Even for the simplest H-like
ions, the rates for the collision process are not sufficiently converged to
match the accuracy of the Hitomi observation.

\section{Charge exchange}

Charge exchange (CX) occurs when an ion collides with a neutral atom, and
catches an electron from the atom. The captured electron often relaxes down to
the ground via line emission. The typical CX cross section at $\sim$keV energies
is two orders of magnitude larger than the electron-impact cross section, making
CX extremely effective for converting neutrals into singly charged ions, and for
reducing the ionization degree of hot plasma.

CX emission shows unique features in the X-rays and UV bands. Since the captured
electron is often caught into a high Rydberg state, it creates transitions at
much higher energies than for collisional excitation, which contributes mostly
to low Rydberg states. CX also differs much from other recombination processes,
because CX capture is quasi-resonant, i.e., strongly selective on Rydberg states
particularly by their principle quantum numbers $n$, while ordinary
recombinations have no such selection. 

\begin{figure*}[!htbp]
\centering
\resizebox{0.6\hsize}{!}{\includegraphics[angle=0]{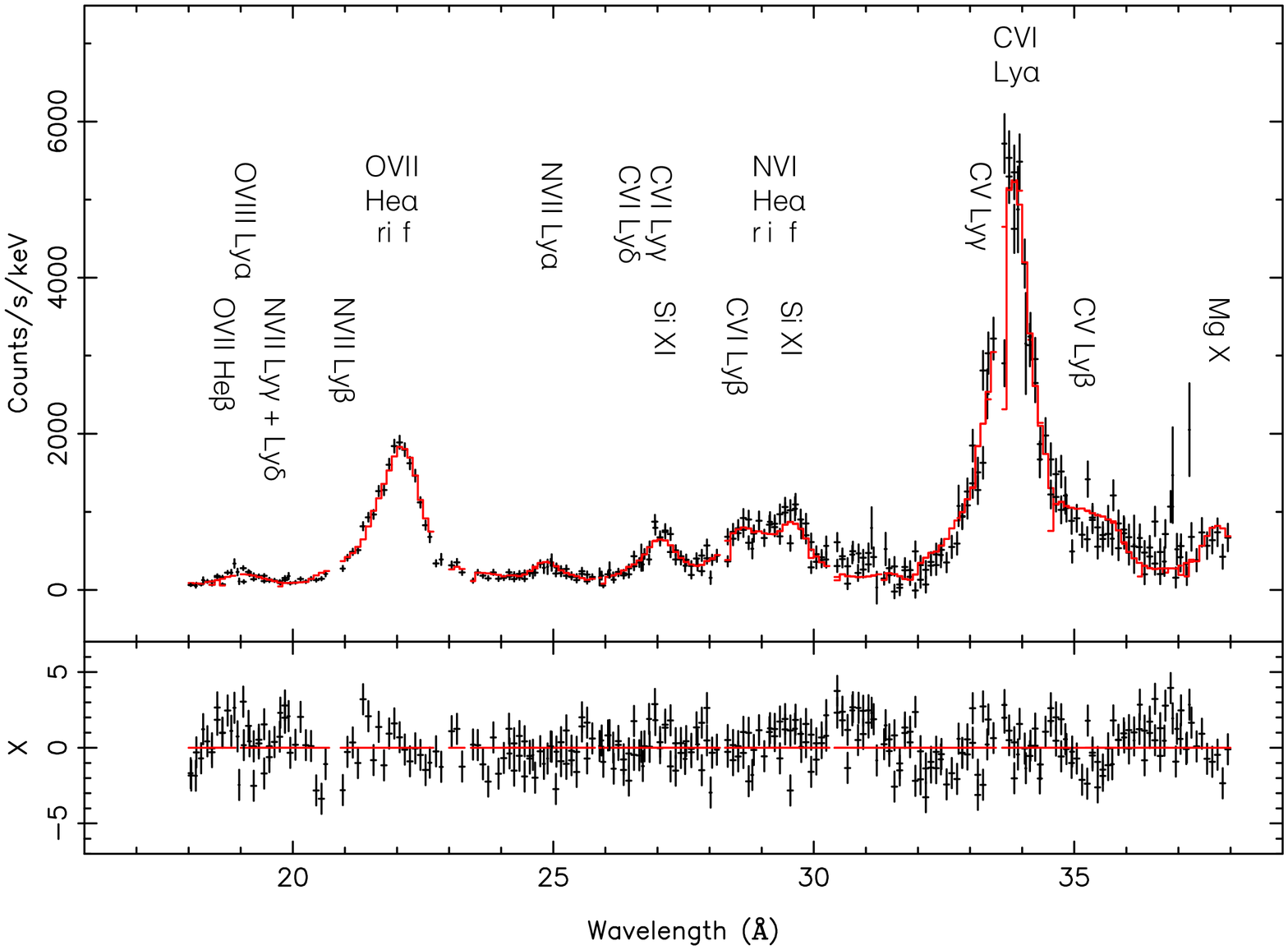}}
\caption{CX fitting (red) to the XMM-Newton RGS spectrum (black) of comet C/2000 WM1 (LINEAR).
\vspace{0.5cm}
}
\label{fig:c2000}
\end{figure*}

For a long time, CX studies suffered the lack of an accurate plasma emission
model. In 2016, we published a ground-breaking plasma code dedicated to the CX
X-ray modeling \cite{gu2016a}. Our model incorporates state-of-the-art atomic
calculations and experimental data. By modeling CX lines for different ions, the
code allows accurate diagnostics of the ionization state, abundances, and
relative velocity of the impinging ions. As shown in Fig.~\ref{fig:c2000}, for
the first time, our model gives a successful fit of the high-resolution X-ray
spectrum from comet C/2000 WM1. We have also explored the CX phenomenon in other
astrophysical objects, such as supernova remnants \cite{gu2016b}, galaxies
\cite{pinto2016}, and AGNs \cite{gu2017}. We can also model CX at lower
energies, e.g., UV, optical, and infrared bands, for interactions between
lowly-ionized ions and atoms/molecules. There is a vast discovery space left to
explore for CX, which may potentially make CX one of the future highlights of
the astronomical research.

So far the most intriguing application of the charge exchange model is the
mysterious 3.5 keV line feature. In 2014, two groups reported to have detected a
weak line feature at $\sim$3.5 keV in the X-ray spectra of clusters of galaxies,
which does not match any known thermal line \cite{boyarsky2014, bulbul2014}. It
immediately vitalized the astrophysical and particle physics communities,
leading to a great amount of follow-up journal papers, most of which discussed
the possibility that this line is created by radiative decay of sterile
neutrinos known as a dark matter candidate. In \cite{gu2015}, we pointed out
that actually a more natural, alternative explanation to the line: CX emission
by hot, fully-ionized sulfur ions colliding with cold, neutral atoms. The two
ingredients, the neutral and ionized particles, co-exist in galaxy clusters. Our
CX model can sufficiently explain the observed strength of the putative feature
at $\sim 3.5$ keV.  Recently, we have confirmed our theoretical calculations by
laboratory measurements using an electron beam ion trap experiment
\cite{shah2016}.

\section{Photoionised outflows in active galactic nuclei}

Supermassive black holes (SMBHs) at the core of active galactic nuclei (AGN)
grow through accretion of matter from their host galaxies. This infall of matter
onto SMBHs is accompanied by {\it outflows of photoionised gas}, which transport
matter and energy away from the nucleus, and thus couple the SMBHs to their host
galaxies and beyond. These outflows from AGN are part of a {\it feedback
mechanism} between the growth of SMBHs and their host galaxies.  It is through
this feedback that AGN shape the galaxy population, and this is now a standard
paradigm in cosmological simulations of galaxy formation.

Ionised outflows are best studied through high-resolution X-ray spectroscopy of
their absorption line spectra, see review\cite{crenshaw2003}. The {\it
XMM-Newton}'s RGS and {\it Chandra}'s LETGS and HETGS have vastly advanced our
knowledge of these outflows in past years.  From spectroscopy of the observed
lines in X-rays, their column density $N_{\rm H}$, ionisation parameter $\xi$,
and velocities are derived. These parameters can be then used to trace the
origin and launching mechanism of outflows in AGN.

The location of the ionised outflows in AGN needs to be established in order to
discriminate between different outflow mechanisms, and to determine their mass
outflow rates and kinetic luminosities, which are essential parameters in
assessing their impact on their surroundings and their contribution to AGN
feedback. The distance of an absorber to the ionising source can be determined
from estimates of the electron density of the absorbing gas, which in turn can
be measured from the ionisation/recombination timescale of the absorber. Such a
study was first done as part of a large multi-wavelength campaign on Seyfert-1
galaxy Mrk 509 \cite{Kaas12}, in which distances of 5--100 pc were derived,
pointing to an origin in the narrow-line region (NLR) or torus region of the
AGN.

In 2013--2014 we carried out an ambitious multi-wavelength campaign on NGC~5548.
This AGN was discovered to be obscured in X-rays with mainly narrow emission
features imprinted on a heavily absorbed continuum. This obscuration is thought
to be caused by a stream of clumpy weakly-ionised gas located at distances of
only light days from the black hole and partially covering the X-ray source and
the BLR. From its associated broad UV absorption lines detected in Hubble Space
Telescope (HST) COS spectra, the obscurer is found to be outflowing with
velocities of up to 5000 km~s$^{-1}$. As the ionising UV/X-ray radiation is
being shielded by the obscurer, new weakly-ionised features of UV and X-ray
absorber outflows have been detected. Compared to normal warm absorber outflows
commonly seen in Seyfert-1s at pc scale distances, the remarkable obscurer in
NGC~5548 is a new breed of weakly-ionised, higher-velocity outflowing gas, which
is much closer to the black hole and extends to the BLR. As reported in
\cite{Kaas14} the outflowing obscurer is likely to originate from the accretion
disk. 

\begin{figure*}[!tbp]
\centering
\includegraphics[width=.70\textwidth]{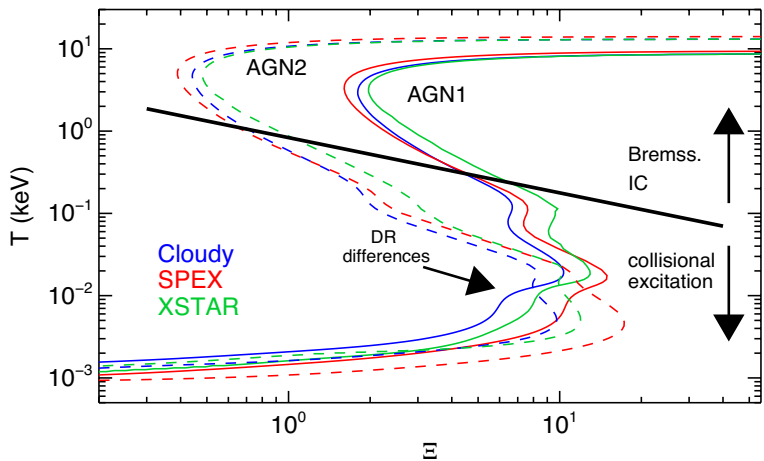}
\caption{Photo-ionisation equilibrium solutions derived by different photo-ionisation codes for unobscured (AGN1) and obscured (AGN2) SEDs of NGC~5548 \cite{Mehd16}; the pressure form of ionisation parameter $\Xi \equiv F/n_{\rm H}ckT$, where $F$ is the ionising flux.}
\label{cartoon_fig}
\end{figure*}

Plasma models and atomic data play a crucial role in diagnosis and
interpretation of astrophysical spectra, thus influencing our understanding of
the universe. Therefore, realistic and accurate models are needed to understand
the data. We carried out a systematic comparison of photo-ionisation codes, to
understand how differences in plasma models and atomic data impact X-ray
spectroscopic studies of outflows in AGN. The modelling uncertainties on the
parameters of photo-ionised plasmas are generally larger than the observational
uncertainties for X-ray bright AGN. In particular, there are substantial
differences between the models at lower temperatures where cooling by
collisional excitation (X-ray line emission) is balanced by photo-ionisation
heating (Fig. \ref{cartoon_fig}). The differences can be attributed to
differences in the excitation rates as well as in the dielectronic recombination
rates, leading to deviations in the predicted ionic column densities by
different models. The results highlight the importance of continuous development
and enhancement of the models and atomic data, which are incorporated in the
photo-ionisation codes. To this end, we have been developing a new
photo-ionisation model for the {\tt SPEX} package, called {\tt pion}.

The {\tt pion} model is a self-consistent photo-ionisation model that calculates
both the ionisation balance and the spectrum, without requiring to run
pre-calculated runs with {\tt XSTAR} or {\tt Cloudy}. The {\tt pion} model is
developed to promptly calculate all the steps of photo-ionisation modelling and
spectral fitting in {\tt SPEX}. It uses the ionising radiation from the
continuum components set by the user in {\tt SPEX}. So during spectral fitting,
as the continuum varies, the ionisation balance and the spectrum of the photo-ionises
plasma are recalculated at each stage. This means while using realistic
broadband continuum components to fit the data, the photo-ionisation balance and
the spectrum are calculated accordingly by the {\tt pion} model. So rather than
assuming an SED shape for ionisation balance calculations, the {\tt pion} model
provides a more accurate approach for determining the intrinsic continuum and
the ionisation balance.

\section{Future instruments}

The X-ray Astronomy Recovery Mission ({\it XARM}) will be the duplicate of the
{\it Hitomi} calorimeter and is scheduled to launch in early 2021. The X-ray
Calorimeter Spectrometer contains 36 pixels, covering a field of view of
3$\times$3 arcmin$^2$ with an effective area of $\sim 200$~cm$^2$. It has a fine
spectral resolution ($\lesssim7$~eV) throughout its entire energy band (0.3-12
keV). The main science goal is to investigate the structure of the Universe and
physics at extreme conditions.

{\it Arcus} has been proposed to NASA as a Medium-Class Explorer in 2016
\cite{smith2016}. It will achieve an unprecedented sensitivity, with an
effective area $\gtrsim 500~{\rm cm^2}$ and a resolving power $\lambda / \Delta
\lambda \gtrsim 3000$ in the 10 to 50 \AA\ wavelength range. It is a significant
improvement over all the existing and approved X-ray spectrometers. {\it Arcus}
will be ready to launch in July 2023, aiming at addressing the formation and
evolution of astronomical objects in different scales from stars to cluster of
galaxies.

\begin{figure}
\resizebox{\hsize}{!}{\includegraphics[angle=0]{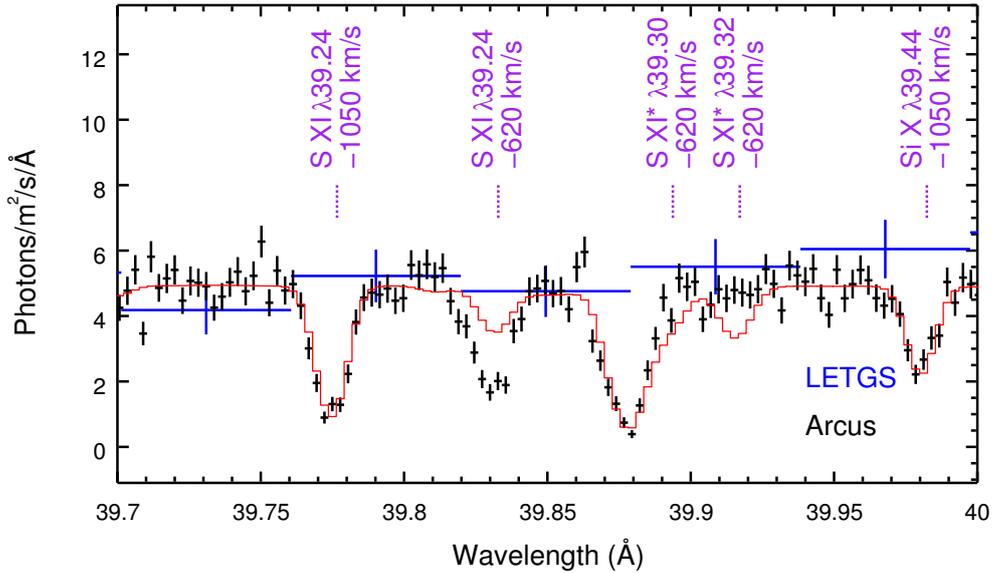}}
\caption{Observed {\it Chandra}/LETGS (in blue, with 345~ks exposure) spectrum of NGC\,5548 in the neighbourhood of S XI $\lambda39.24$ (in the observed frame) and a 300~ks simulation of the same spectrum with {\it Arcus} (in black). Two outflow components with velocities of $-620~{\rm km~s^{-1}}$ and $-1050~{\rm km~s^{-1}}$ are present in the wavelength range. The red solid line is a model with density of $n_{\rm H}=10^{16}~{\rm m^{-3}}$, while the simulated {\it Arcus} data correspond to $n_{\rm H}=10^{6}~{\rm m^{-3}}$. Under certain conditions, the metastable ($\ast$) to ground absorption line ratio might vary with the density of the photoionised plasma.}
\label{fig:arcus}
\end{figure}

Fig.~\ref{fig:arcus} demonstrates the advantage of {\it Arcus} in terms of
statistics and spectral resolution. Such high quality spectra can be used to
constrain the density of ionised outflows in AGN via the metastable to ground
absorption line ratio.

The {\it ATHENA} X-ray observatory will be the next large ESA mission to study
the hot and energetic universe and is planned for launch in 2028
\cite{nandra2013}. The observatory features a Wide-Field Imager (WFI) X-ray
camera and an X-ray Integral Field Unit (X-IFU). The X-IFU consists of a large
array of Transition-Edge Sensors (TES), which enable spatially resolved
high-resolution spectroscopy with a spectral resolution of 2.5 eV in the
$\sim$0.3-12 keV band and an effective area of 1--2~m$^{2}$. The main goal of
the observatory is to learn more about the formation of the large-scale
structure of the universe and the role of black holes in this formation process.
For example, using X-IFU, turbulence in the hot gas of massive clusters of
galaxies can be measured at a much higher spatial resolution of 5 arcseconds. In
nearby clusters, this will allow us to spatially resolve the turbulence induced
by giant bubbles of gas blown by a central super-massive black hole. By pointing
ATHENA to bright and very distant gamma-ray bursts within a few hours, X-IFU
will be used to probe the teneous Warm-Hot Intergalactic Medium (WHIM) that
traces the connecting threads of the cosmic web in absorption. With the X-IFU
instrument, ATHENA promises to become a very powerful mission for
high-resolution X-ray spectroscopy of especially spatially extended sources.

\acknowledgments

SRON is supported financially by NWO, the Netherlands Organization for
Scientific Research. 


\end{document}